\documentclass [12pt,a4paper]{article}
\usepackage{bbm}
\usepackage{amsfonts}
\usepackage{mathrsfs}
\usepackage{amsthm}
\usepackage{latexsym}
\usepackage{amssymb}
\usepackage{amsmath}
\usepackage{color}
\usepackage{dcolumn}
\usepackage{extarrows}
\usepackage{hyperref}
\makeatletter
 \@addtoreset{equation}{section}
 
\makeatother

\usepackage{lastpage}
\usepackage{epsfig}

\begin{document}
\begin{center}
\textbf{\Large{Mutually unbiased maximally entangled bases in  $\mathbb{C}^d\otimes\mathbb{C}^d$ } }\footnote { The work of J. Liu and M. Yang were supported by the National Natural Science Foundation of China(NSFC) under Grant 61379139, 11526215 and the strategic Priority Research Program of the Chinese Academy of Sciences under Grant XDA06010701. The work of K. Feng was supported by the NSFC under Grant 11471178, 11571007 and the Tsinghua National Lab. for Information Science and Technology. \\
J. Liu and M. Yang are with the State Key Laboratory of Information
Security, Institute of Information Engineering, Chinese Academy
of Sciences, Beijing 100093, China(e-mails: liujunying@iie.ac.cn; yangminghui6688@163.com)\\
K. Feng is with the Department of Mathematical Sciences, Tsinghua University, Beijing, 100084, China(e-mail: kfeng@math.tsinghua.edu.cn)}

\end{center}

\begin{center}
\small Junying Liu, Minghui Yang,  Keqin Feng
\end{center}

%-------------------------------------------------------------------------

\noindent\textbf{Abstract} We study mutually unbiased maximally entangled bases (MUMEB's) in bipartite system  $\mathbb{C}^d\otimes\mathbb{C}^d (d \geq 3)$. We generalize the method to construct MUMEB's given in  [16], by using
any commutative ring $R$ with $d$ elements and generic character of $(R,+)$ instead of $\mathbb{Z}_d=\mathbb{Z}/d\mathbb{Z}$.
Particularly, if $d=p_1^{a_1}p_2^{a_2}\ldots p_s^{a_s}$ where $p_1, \ldots,  p_s$ are distinct primes and $3\leq p_1^{a_1}\leq\cdots\leq p_s^{a_s}$, we present $p_1^{a_1}-1$ MUMEB's in $\mathbb{C}^d\otimes\mathbb{C}^d$ by taking $R=\mathbb{F}_{p_1^{a_1}}\oplus\cdots\oplus\mathbb{F}_{p_s^{a_s}}$, direct sum of finite fields (Theorem 3.3).


\section{Introduction}
 \noindent Mutually unbiased bases (MUB) play central roles in quantum kinematics [1], quantum state tomography [2,3] and quantifying wave-particle duality in multipath interferometers [4]. Moreover, the importance of the MUB has been demonstrated in various tasks of quantum information processing such as quantum key distribution [5], cryptographic protocols [5,6], mean king problem [7] and quantum teleportation and superdense coding [8-10].

\textbf{Definition 1.1} Two orthogonal bases $\mathcal{B}_1=\{|\phi_i\rangle: 1\leq i\leq d\}$ and $\mathcal{B}_2=\{|\psi_i\rangle: 1\leq i\leq d\}$ of $\mathbb{C}^d$ are called mutually unbiased if
$$|\langle \phi_i|\psi_j\rangle|=\frac{1}{\sqrt{d}},\ \ \ \ (1\leq i, j\leq d).$$
A set of orthonormal bases $\mathcal{B}_1, \mathcal{B}_2, \ldots, \mathcal{B}_m$ in $\mathbb{C}_d$ is called mutually unbiased bases (MUB) if every pair of $\mathcal{B}_i$ and $\mathcal{B}_j$ $(1\leq i\neq j\leq d)$ is mutually unbiased.

For each integer $d\geq 2$, let $N(d)$ be the maximal number $m$ such that there exist MUB $\{\mathcal{B}_1, \mathcal{B}_2, \ldots, \mathcal{B}_m\}$ in $\mathbb{C}^d$. It is proved that $N(d)\leq d+1$ and $N(d)=d+1$ if $d$ is a prime power. The last result has been
proved in different points of view and one of proofs was given in [11] by using finite field (for $d=p^m, p\geq 3$) and Galois ring
$GR(4,m)$ (for $d=2^m$). If $d$ is not a prime power, to determine the value of $N(d)$ is an open problem.

In this paper we study mutually unbiased maximally entangled bases (MUMEB's) in bipartite system $\mathbb{C}^d\otimes\mathbb{C}^{d}$.
For $2\leq d\leq d'$, a (pure) maximally entangled state in  $\mathbb{C}^d\otimes\mathbb{C}^{d'}$ can be written as
$$|\psi\rangle=\frac{1}{\sqrt{d}}\sum_{i=0}^{d-1}|e_i\rangle\otimes|e'_i\rangle,$$
where $\{|e_i\rangle: 0\leq i\leq d-1\}$ and $\{|e'_i\rangle: 0\leq i\leq d'-1\}$ are orthonomal bases of $\mathbb{C}^d$ and
$\mathbb{C}^{d'}$ respectively. Maximally entangled states play vital role in quantum information processing [1, 12-15]. Let $M(d, d')$ be the maximal size $m$ of mutually unbiased maximally entangled bases (MUMEB's) $\{\mathcal{B}_1, \mathcal{B}_2, \ldots, \mathcal{B}_m\}$ in  $\mathbb{C}^d\otimes\mathbb{C}^{d'}$, where each $\mathcal{B}_i$ is an orthonormal basis of $\mathbb{C}^d\otimes\mathbb{C}^{d'}$ consisted of $dd'$ maximally entangled states and for $1\leq i \neq j\leq m$, $\mathcal{B}_i$
 and $\mathcal{B}_j$ are mutually unbiased. One of the basic problem is how large of $M(d,d')$ could be. A general method to construct MUMEB in $\mathbb{C}^d\otimes\mathbb{C}^{d'}$ for $d'=kd (k\geq 1)$ has been given in [16] and showed $M(2, 4)\geq 5$ and $M(2, 6)\geq 3$. Namely, five and three MUMEB's have been constructed in $\mathbb{C}^2\otimes\mathbb{C}^4$ and  $\mathbb{C}^2\otimes\mathbb{C}^6$ respectively by using this construction method. In [17], authors presented a method to construct a pair of MUMEB's in $\mathbb{C}^d\otimes\mathbb{C}^{2^ld'}$ for all $l\geq1$ from a pair of MUMEB's in $\mathbb{C}^d\otimes\mathbb{C}^{d'}$.

 In this paper we study MUMEB's in $\mathbb{C}^d\otimes\mathbb{C}^{d} (d\geq 2)$. Firstly, in Section 2 we slightly generalize the construction method presented in [16] by using any commutative ring $R$ with $d$ elements and generic character of $(R, +)$ instead of $\mathbb{Z}_d=\mathbb{Z}/d\mathbb{Z}$. Then, in Section 3, we construct MUMEB's in  $\mathbb{C}^d\otimes\mathbb{C}^{d}$ by using this generalization (Theorem 3.2). Particularly, if $d=p_1^{a_1}\ldots p_s^{a_s}$ where $p_1,\ldots,p_s$ are distinct primes and $3\leq p_1^{a_1}\leq\cdots\leq p_s^{a_s}$, we get $p_1^{a_1}-1$ MUMEB's in $\mathbb{C}^d\otimes\mathbb{C}^{d}$ by taking $R=\mathbb{F}_{p_1^{a_1}}\oplus\cdots\oplus\mathbb{F}_{p_s^{a_s}}$. In Section 4 we give conclusion and raise some open problems.

\section{General Construction on MUMEB's in $\mathbb{C}^d\otimes\mathbb{C}^{d} (d\geq 2)$ }

\ \ \ \  In this section we introduce the general construction on MUMEB's in $\mathbb{C}^d\otimes\mathbb{C}^{d}$ given in [16] with a slight generalization. Namely we use any commutative ring $R$ with $d$ elements and generic additive character instead of  $\mathbb{Z}_d$.

We fix an orthonormal basis $\{e_r: r\in R\}$ of $\mathbb{C}^d$ and consider the following maximally entangled state
\begin{equation}
|\psi_U\rangle=\frac{1}{\sqrt{d}}\sum_{r\in R}|e_r\rangle\otimes U|e_r\rangle
\end{equation}
where $U$ is an unitary operator (matrix) of $\mathbb{C}^d$ so that $\{U|e_{r}: r\in R\}$ is an orthonormal basis of $\mathbb{C}^{d}$. Let $U=(u_{r,s})_{r,s\in R}(u_{rs}\in\mathbb{C})$, then
$$U|e_r\rangle=\sum_{s\in R}u_{r,s}|e_s\rangle\ \ (r\in R).$$
Remark that $U$ is unitary if and if $U^fU=I_d$ where $U^f=(u_{rs}^f), u_{rs}^f=\overline{u}_{sr}$. Therefore $U$ is unitary if and only if for any $r, s\in R$,
\begin{displaymath}
 \sum_{\substack{l\in R}}\overline{u_{lr}}{u_{ls}}=\delta_{rs}
 = \left\{ \begin{array}{ll}
1, & \textrm{if $r=s$}\\
0, & \textrm{otherwise.}
\end{array} \right.
\end{displaymath}

A character of the additive group $(R,+)$ is an isomorphism of groups $\lambda: (R,+)\rightarrow\langle\zeta_d\rangle (\zeta_d=e^{\frac{2\pi\sqrt{-1}}{d}})$ which means that $\lambda(r+s)=\lambda(r)\lambda(s)$, $\lambda(0)=1$, and
$\overline{\lambda}(r)=\lambda^{-1}(r)=\lambda(-r).$ In this paper, we assume that \\
 there exists a ``generic" character $\lambda$ of $(R,+)$ which means that for any $0\neq a\in R$,
$$\sum_{r\in R}\lambda(ar)=0.$$
Next we act on $|\psi_U\rangle$ (of (2.1)) by Pauli (or called Weyl-Heisenberg) operators $H_{\xi,\eta} (\xi,\eta\in R)$ to get the following $d^2$ maximally entangled states
\begin{equation}
H_{\xi,\eta}|\psi_U\rangle=\frac{1}{\sqrt{d}}\sum_{r\in R}\lambda(r\xi)|e_r\rangle\otimes U|e_{r+\eta}\rangle,
\end{equation}
where $\lambda$ is a fixed generic character of $(R,+)$.

\textbf{Lemma 2.1} (1). For any unitary operator $U$ on $\mathbb{C}^{d}$,
\begin{equation}
\Phi_U=\{H_{\xi,\eta}|\Psi_U\rangle: \xi,\eta\in R\}
\end{equation}
is an orthonomal maximally entangled basis (MEB) in $\mathbb{C}^d\otimes\mathbb{C}^{d}$.

(2). For two unitary operators $U$ and $V$ on $\mathbb{C}^{d}$, let $W=U^fV=(\omega_{rs})_{r,s\in R}.$ Then two MEB's $\Phi_U$ and
$\Phi_V$ in $\mathbb{C}^d\otimes\mathbb{C}^{d}$ are mutually unbiased if and only if for any $\xi,\eta\in R$,
$$|\sum_{r\in R}\lambda(\xi r)\omega_{r,r+\eta}|=1.$$
 \begin{proof}(1). For $\xi,\xi',\eta,\eta'\in R$, the hermitian inner product of $H_{\xi,\eta}|\psi_U\rangle$ and
$H_{\xi',\eta'}|\psi_U\rangle$ is, by (2.2)
\begin{eqnarray*}
&&\frac{1}{d}\sum_{r,r'\in R}\overline{\lambda}(r\xi)\lambda(r'\xi')\langle e_r|e_{r'}\rangle\langle e_{r+\eta}|U^fU| e_{r'+\eta'}\rangle \\
&&=\frac{1}{d}\sum_{r\in R}\lambda((r(\xi'-\xi))\langle e_{r+\eta}|e_{r+\eta'}\rangle \ \ \ (\textrm{since}\ \langle e_r|e_{r'}\rangle =\delta_{r,r'}\ \textrm{and}\ U^fU=I_d ) \\
&&=\frac{1}{d}\sum_{r\in R}\lambda((r(\xi'-\xi))\delta_{\eta,\eta'} =\delta_{\xi,\xi'}\cdot\delta_{\eta,\eta'}\ (\textrm{since}\ \lambda \ \textrm{is} \ \textrm{generic} )
\end{eqnarray*}
Therefore $\Phi_U$ is an orthonomal MEB of $\mathbb{C}^d\otimes\mathbb{C}^{d}$.

(2) By Definition (1.1), $\Phi_U$ and $\Phi_V$ are mutually unbiased if and only if for any $\xi,\xi',\eta,\eta'\in R,$
$$|\langle\psi_U|H_{\xi,\eta}^f H_{\xi',\eta'}|\psi_V\rangle|=1/d.$$
In fact, by (2.2) we have
\begin{equation*}\begin{split}
\langle \psi_U|H_{\xi,\eta}^fH_{\xi'\eta'}|\psi_V\rangle& =\frac{1}{d}\sum_{r,r'\in R}\overline{\lambda}(r\xi)\lambda(r'\xi')
                                                            \langle e_r|e_{r'}\rangle\langle e_{r+\eta}|U^fV|e_{r'+\eta'}\rangle\\
                                  &= \frac{1}{d}\sum_{r\in R}\lambda(r(\xi'-\xi))\langle e_{r+\eta}|W|e_{r+\eta'}\rangle\\
                                  &= \frac{1}{d}\sum_{r\in R}\lambda(r(\xi'-\xi))\omega_{r+\eta',r+\eta} (\textrm{let} \ l=r+\eta')\\
                                  &=\frac{1}{d}\overline{\lambda}(\eta(\xi'-\xi))\sum_{l\in R}\lambda(l(\xi'-\xi))w_{l,l+\eta-\eta'}.
                                  \end{split}\end{equation*}
Since $|\overline{\lambda}(\eta(\xi'-\xi))|=1$, we know that $\Phi_U$ and $\Phi_V$ are mutually unbiased if and only if for any $\xi,\eta\in R$, $|\sum_{l\in R}\lambda(l\xi)\omega_{l,l+\eta}|=1$.
\end{proof}
In the next section we will find specific unitary $U_i(1\leq i\leq m)$ for some $m$, such that $\Phi_{U_i}(1\leq i\leq m)$ are mutually unbiased by using the criterion given in Lemma 2.1 (2).

\section{Construction of MUMEB's in $\mathbb{C}^d\otimes\mathbb{C}^{d}$ }

Let $d\geq 2$, $R$ be a commutative ring with $d$ elements and generic additive character $\lambda$, $R^{\ast}$ be the group of
invertible elements of $R$. For each $b\in R^{\ast}$ we define an operator $U^{(b)}$ on $\mathbb{C}^{d}$ by
\begin{equation}
U^{(b)}=(u_{rs}^{(b)})_{r,s\in R}, \ u_{rs}^{(b)}=\delta_{br,s}
\end{equation}
Namely, for each $r\in R$,
\begin{equation}
U^{(b)}|e_r\rangle=\sum_{l\in R}u_{rl}^{(b)}|e_l\rangle=\sum_{l\in R}\delta_{br,l}|e_l\rangle=e_{br}\rangle
\end{equation}
which means that $U^{(b)}$ is a permutation matrix. From (3.2) we know that for $a, b\in R^{\ast}$, $U^{(a)}=I_d$ if
and only if $a=1$, and
$$U^{(a)} U^{(b)}=U^{(ab)},{U^{(b)}}^f=U^{(b^{-1})}={U^{(b)}}^{-1}.$$
Therefore ${U^{(b)}}^fU^{(b)}=I_d.$ Namely, $U^{(b)}$ is unitary for all $b\in{R^{\ast}}.$

\textbf{Lemma 3.1} If $a,b\in R^{\ast}$ and $a-b\in R^{\ast}$, two orthonomal MEB's $\Phi_{U^{(a)}}$ and $\Phi_{U^{(b)}}$ are
mutually unbiased.
\begin{proof}
 By Lemma 2.1(2), we need to
show that for any $\xi,\eta\in R$,
$$|\sum_{r\in R}\lambda(\xi r)u_{r,r+\eta}^{(c)}|=1,$$
where $c=ba^{-1}$ since ${U^{(a)}}^fU^{(b)}=U^{(c)}$. By (3.1),
\begin{equation}
\sum_{r\in R} \lambda(\xi r){u^{(c)}_{r,r+\eta}}=\sum_{r\in R}\lambda(\xi r)\delta_{cr,r+\eta}
\end{equation}
The assumption $a-b\in R^{\ast}$ implies that $c-1=-a^{-1}(a-b)\in R^{\ast}.$ Thus for any $\eta\in R$ the
equation $cr=r+\eta$ has unique solution $r=\eta(c-1)^{-1}\in R$. Therefore
$$|\sum_{r\in R}\lambda(\xi r)u_{r,r+\eta}^{(c)}=|\lambda(\xi\eta(c-1)^{-1})|=1$$
which means that $\Phi_{U^{(a)}}$ and $\Phi_{U^{(b)}}$ are mutually unbiased.
\end{proof}

As a direct consequence of Lemma 3.1, we get the following result.

\textbf{Theorem 3.2} Let $d\geq 3$, $R$ be a commutative ring with $d$ elements and generic character $\lambda$ of $(R,+)$, $R^{\ast}$ be the group
of invertible elements of $R$. If there exists a subset $S$ of $R^{\ast}$, $|S|=m\geq 2$, satisfying the following condition

$(\ast)$ For any distinct elements $b,b'$ in $S, b-b' \in R^{\ast}$.\\
Then there exist $m$ MUMEB's in $\mathbb{C}^d\otimes\mathbb{C}^d$.
\begin{proof} Let $S=\{b_1,\ldots,b_m\}$. $U^{(b_i)}(1\leq i\leq m)$ are unitary operators on  $\mathbb{C}^d$
defined by (3.1). Then $\Phi_{U^{(b_i)}}=\{H_{\xi,\eta}|\Phi_{U^{(b_i)}}\rangle: \xi,\eta\in R\} (1\leq i\leq m),$
defined by (2.2), are orthonomal maximally entangled bases in $\mathbb{C}^d\otimes\mathbb{C}^d$ (Lemma 2.1). From
assumption $(\ast)$ and Lemma 3.1 we know that these $m$ MEB's are mutually unbiased. This completes the proof of Theorem 3.2.
\end{proof}
As an application of Theorem 3.2, we have the following result.

\textbf{Theorem 3.3} Let $d=p_1^{a_1}p_2^{a_2}\ldots p_s^{a_s}, 3\leq p_1^{a_1}\leq p_2^{a_2}\leq\cdots\leq p_s^{a_s}$, where $p_1,\ldots,p_s$ are distinct primes. Then $M(d,d)\geq p_1^{a_1}-1.$ Namely, there exist $p_1^{a_1}-1(\geq 2)$ MUMEB's in $\mathbb{C}^d\otimes \mathbb{C}^d$.
\begin{proof}
Let $q_i=p_i^{a_i}(\geq 3)$. We take $R=\mathbb{F}_{q_1}\oplus\mathbb{F}_{q_2}\oplus\cdots\oplus\mathbb{F}_{q_s}$ (a
direct sum of finite fields). For each $i$ we have the trace map
$$T_i: \mathbb{F}_{q_i}\rightarrow \mathbb{F}_{p_i}, T_i(x)=x+x^{p_i}+x^{{p_i}^2}+\cdots+x^{{p_i}^{a_i-1}}(x \in \mathbb{F}_{q_i})$$
It is known that $\lambda_i:\mathbb{F}_{q^i}\rightarrow \langle \zeta_{p_i}\rangle$, $\lambda_i(x)=\zeta_{p_i}^{T_i(x)}$ is a generic character of $(\mathbb{F}_{q_i}, +).$ Namely, for any $a\in \mathbb{F}_{q_i}^{\ast}=\mathbb{F}_{q_i}\backslash \{0\}$, $\sum_{x\in\mathbb{F}_{q_i}}\lambda_i(ax)=\sum_{x\in\mathbb{F}_{q_i}}\lambda(x)_i=0$.\\
For $x=(x_1,\ldots,x_s)\in R \ (x_i\in \mathbb{F}_{q_i}), $ we define
$$\lambda(x)=\lambda_1(x_1)\ldots\lambda_s(x_s)$$
which is a character of $(R,+)$. Moreover, if $0\neq c=(c_1,\ldots,c_s)\in R,$ there exists $j (1\leq j\leq s)$ such that
$c_j\neq 0$. Then $cx=(c_1x_1,\ldots,c_sx_s)$ and
$$\sum_{x\in R}\lambda(cx)=\sum_{\substack{x_i\in {\mathbb{F}_{q_i}}\\(1\leq i\leq s)}}\lambda_1(c_1x_1)\cdots\lambda_s(c_sx_s)
=\prod_{i=1}^s\sum_{x_i\in\mathbb{F}_{q_i}}\lambda_i(c_ix_i)=0$$
since $\sum_{x_j\in\mathbb{F}_{q_j}}\lambda_j(c_jx_j)=0.$ Therefore the character $\lambda$ is generic.

Let $\mathbb{F}_{q_1}^{\ast}=\mathbb{F}_{q_1}\backslash\{0\}=\{b_1^{(1)},\ldots,b_{q_1-1}^{(1)}\}$. By $q_i-1\geq q_1-1 (2\leq i\leq s)$, we can take a subset $\{b_1^{(i)},\ldots,b_{q_1-1}^{(i)}\}$ of $\mathbb{F}_{q_i}^{\ast} (2\leq i\leq s)$. Then
$$b_l=(b_l^{(1)},b_l^{(2)}\ldots,b_l^{(s)})\in \mathbb{F}_{q_1}^{\ast}\oplus\mathbb{F}_{q_2}^{\ast}\oplus\cdots\oplus\mathbb{F}_{q_s}^{\ast}=R^{\ast} \ (1\leq l\leq q_1-1).$$
Moreover, for $1\leq l\neq l'\leq q_1-1,$ we have $b_l^{(i)}\neq b_{l'}^{(i)}\ (1\leq i\leq s).$ Therefore
$$b_l-b_{l'}=(b_l^{(1)}-b_{l'}^{(1)},\ldots,b_l^{(s)}-b_{l'}^{(s)})\in R^{\ast}$$
which means the subset $S=\{b_1,\ldots,b_{q_1-1}\}$ of $R^{\ast}$ satisfies the assumption $(\ast)$ of Theorem 3.2. By Theorem 3.2,
there exist $|S|=p_1^{a_1}-1$ MUMEB's in $\mathbb{C}^d\otimes\mathbb{C}^d$.
\end{proof}

\section{Conclusion }

\ \ \ \ We slightly generalize the method to construct mutually unbiased (orthonormal) maximal entangled bases (MUMEB's) in  $\mathbb{C}^d\otimes\mathbb{C}^d$ given in [16] by using arbitrary commutative ring $R$ with $d$ elements and generic additive character instead of $\mathbb{Z}_d$. For $d=q_1\ldots q_s$, where $3\leq q_1\leq\cdots\leq q_s$ and $q_i=p_i^{a_i}$ where
$p_1,\ldots,p_s$ are distinct primes, we present $q_1-1$ MUMEB's in $\mathbb{C}^d\otimes\mathbb{C}^d$ by taking $R$ as the direct product of finite fields $\mathbb{F}_{q_1}\oplus\cdots\oplus\mathbb{F}_{q_s}$. Therefore $M(d,d)\geq q_1-1$ where $M(d,d')$ is the maximal size of MUMEB's in $\mathbb{C}^d\otimes\mathbb{C}^{d'}$. Particularly, if $d=p^a\geq 3$ is a power of prime number $p$, then
 $M(d,d)\geq d-1.$

 Let $N(d)$ be the maximal size of MUB's in $\mathbb{C}_d$, as we stated in Section 1, $N(d)\leq d+1$ by using Welch bound in sphere
 design theory and $N(d)=d+1$ for $d$ being a power of a prime. For MUMEB case, we raise the following open problems.

 (1). What is a reasonable upper bound of $M(d,d')$? In the first step, can we determine the exact values $M(d,d)$ for smaller $d$
 or $d=p^a$?

 (2). Can we find a method to improve the lower bound of $M(d,d)$ given by Theorem 3.3? More generally, for $d'\geq d$, can we find a
systematic way to construct MUMEB's with large size in $\mathbb{C}^d\otimes\mathbb{C}^{d'}$ ?

\end{document}